# Transformation-optics generalization of tunnelling effects in bi-layers made of paired pseudo-epsilon-negative/mu-negative media

G Castaldi [1], I Gallina [1], V Galdi [1], A Alù [2], and N Engheta [3]

[1] Waves Group, Department of Engineering, University of Sannio, I-82100, Benevento, Italy

[2] Department of Electrical and Computer Engineering, The University of Texas at Austin, Austin, TX 78712, USA

[3] Department of Electrical and Systems Engineering, University of Pennsylvania, Philadelphia, PA 19104, USA

E-mail: vgaldi@unisannio.it

**Abstract.** Transformation-media designed by standard transformation-optics (TO) approaches, based on *real-valued* coordinate-mapping, cannot exhibit *single-negative* (SNG) character unless such character is already possessed by the domain that is being transformed. In this paper, we show that, for a given field polarisation, pseudo-SNG transformation media can be obtained by transforming a domain featuring *double positive* (or *double-negative*) character, via *complex* analytic continuation of the coordinate transformation rules. Moreover, we apply this concept to the TO-based interpretation of phenomena analogous to the tunnelling effects observable in bi-layers made of *complementary* epsilon-negative (ENG) and mu-negative (MNG) media, and explore their possible TO-inspired extensions and generalizations.

## 1. Introduction and background

Transformation optics (TO) (see e.g., [1-5]) is a powerful and systematic approach to the synthesis of metamaterials with broad field-manipulation capabilities. Its technological viability is rapidly getting established, in view of the formidable advances in the fabrication of artificial materials and metamaterials with controllable anisotropy and spatial inhomogeneity properties. The reader is referred to [6-14] for a sparse sampling of recent representative applications, ranging from the celebrated "invisibility cloaking" [7], to hyper/super-lensing scenarios [8-11], and electromagnetic (EM) analogous of relativistic effects [12] and celestial-mechanical phenomena [13-14].

Standard TO is based on the design of a suitable *real-valued* coordinate transformation, which produces the desired field-behaviour in a *curved-metric fictitious* space. Thanks to the formal invariance of Maxwell equations, this behaviour may be translated to a *flat-metric physical* space filled up with an anisotropic, spatially-inhomogeneous "transformation medium" [1-5].

The constitutive parameters of such transformation medium are systematically derived from those of the fictitious space via the Jacobian matrix of the transformation [see, e.g., equation (1) below], and this imposes some restrictions on the material properties that can be obtained by transforming a given scenario into another. For instance, in the standard TO approach, the reciprocity character of the domain that is being transformed cannot be changed [15]. Moreover, the signs of the constitutive-tensor eigenvalues can either be all preserved or all flipped [15]. Thus, e.g., transforming a domain featuring a *double positive* (DPS) character, one can only obtain a DPS or a *double negative* (DNG) transformation medium (and vice-versa).

The latter case is obtained via *coordinate folding*, and allows (though with some caveats [16]) the TO interpretation of interesting effects such as, e.g., superlensing [10-11] and anti-cloaking [17-18].

To sum up, important material properties, such as *non-reciprocity* and *single-negative* (SNG) character, cannot be generated via standard TO, unless already possessed by the domain that is being transformed. In order to overcome these limitations, some possible generalizations have been proposed. For instance, in [15] an extension based on "triple spacetime metamaterials" has been proposed, which allows geometrical interpretation of *non-reciprocal* and *indefinite* media.

In [19], as a possible extension of a general class of transformation slabs, we speculated that, for a given field polarisation, SNG transformation media could be obtained by transforming a domain featuring DPS (or DNG) character, via *complex* analytic continuation of the coordinate transformation. In this paper, we further elaborate on this concept, and address its application to the TO interpretation of the tunnelling effects observed in [20] in connection with bi-layers of (homogeneous, isotropic) epsilon-negative (ENG) and mu-negative (MNG) slabs, and their further generalization to (inhomogeneous, anisotropic) complementary media [21,22]. Our results reproduce as special cases those in [20-22], and suggest other TO-inspired broad tunnelling conditions.

Accordingly, the rest of the paper is laid out as follows. In Section 2, we introduce the problem scenario and formulation. In Section 3, we outline the main analytical derivations, starting from the field calculation, and proceeding with the study of total-transmission conditions. In Section 4, we provide a physical interpretation of the results, and illustrate their relationship with those in [20-21]. In Section 5, we illustrate some novel representative results, not amenable to those in [20-21]. Finally, in Section 6, we provide some brief conclusions and hints to future research directions related to these findings.

## 2. Problem formulation

In [19], in connection with a general class of (DPS or DNG) transformation slabs, we have theoretically speculated that SNG transformation media could be generated by transforming a domain featuring DPS or DNG character, via a suitable analytical continuation of the coordinate transformation in the complex plane. From a physical viewpoint, one intuitively expects such coordinate transformation to exhibit an in-plane *purely imaginary* character, in order to map a *propagating* field solution in the fictitious "above cut-off" space into an *evanescent* "below cut-off" one in the transformed (SNG) domain. In this framework, while a *single* SNG transformation slab would be of limited interest (being inherently opaque), it may be of interest to verify that *paired* (ENG-MNG) transformation-media configurations may support interesting tunnelling phenomena under proper *matching* conditions, in analogy to what was shown in [20-21]. Accordingly, as illustrated in figure 1(a), we consider a *transformation bi-layer* which occupies the region $|x|<d$ in a physical space $(x,y,z)$, characterised by the relative permittivity and permeability tensors

$$\underline{\underline{\varepsilon}}_\alpha(x,y) = \underline{\underline{\mu}}_\alpha(x,y) = \det\left[\underline{\underline{J}}_\alpha(x,y)\right] \underline{\underline{J}}_\alpha^{-1}(x,y) \cdot \left[\underline{\underline{J}}_\alpha^{-1}(x,y)\right]^T, \qquad (1)$$

where $\alpha=1$ for $-d<x<0$, $\alpha=2$ for $0<x<d$, the apex $^T$ denotes matrix transposition, and $\underline{\underline{J}}_\alpha(x,y)$ are the Jacobian matrices of the 2-D coordinate transformations

$$\begin{cases} x' = ia_\alpha u_\alpha(x), \\ y' = i\dfrac{y}{\dot{u}_\alpha(x)} + iv_\alpha(x), \\ z' = z, \end{cases} \qquad (2)$$

from a fictitious (vacuum) space $(x',y',z')$. In (2), and henceforth, $a_\alpha \neq 0$ are real scaling parameters, $u_\alpha(x)$ and $v_\alpha(x)$ are arbitrary continuous real functions, an overdot denotes differentiation with respect to the argument, and a time-harmonic $\exp(-i\omega t)$ dependence is assumed for all field quantities. Moreover, we assume that the derivatives $\dot{u}_\alpha(x)$ are *continuous* and *positive* within the bi-layer region $|x|<d$ [so as to avoid singularities in (2)]. We point out that the transformation in (2) is neither the only one, nor the most general in order to achieve the sought SNG character, and that it was chosen in view of its close resemblance to the (real) transformation considered in [19], so as to exploit (with some minor variations) our previous analytical derivations. As shown in [19], the constitutive tensors in (1) are *real* and *symmetric*, and hence they can be diagonalised, thereby providing a more insightful interpretation. Accordingly, by indicating with $(\xi_\alpha, \upsilon_\alpha, z)$ the principal reference systems (constituted by the orthogonal eigenvectors), and with $\Lambda_{\xi_\alpha}$ and $\Lambda_{\upsilon_\alpha}$ the in-plane eigenvalues, the diagonalised forms can be written as (see [19] for details)

$$\underline{\underline{\tilde{\varepsilon}}}_\alpha(x,y) = \underline{\underline{\tilde{\mu}}}_\alpha(x,y) = \begin{bmatrix} \Lambda_{\xi_\alpha}(x,y) & 0 & 0 \\ 0 & \Lambda_{\upsilon_\alpha}(x,y) & 0 \\ 0 & 0 & -a_\alpha \end{bmatrix}, \qquad (3)$$

where

$$\text{sgn}\left[\Lambda_{\xi_\alpha}(x,y)\right] = \text{sgn}\left[\Lambda_{\upsilon_\alpha}(x,y)\right] = \text{sgn}(a_\alpha). \qquad (4)$$

Therefore, depending on field polarisation and on the sign of the scaling parameters $a_\alpha$, the corresponding transformation media may *effectively* behave as either ENG or MNG, as compactly summarised in table 1. However, since they exhibit by construction $\underline{\underline{\varepsilon}}_\alpha = \underline{\underline{\mu}}_\alpha$, the terms "ENG," "MNG," and "SNG" may appear not entirely appropriate, and throughout the paper we shall refer to them as "pseudo-ENG" (P-ENG), "pseudo-MNG" (P-MNG), and "pseudo-SNG" (P-SNG).
In what follows, generalising the studies in [20-21], we analyse the transmission properties of bi-layers made of paired P-ENG/P-MNG transformation media.

### 3. Analytical derivations

*3.1 Field calculation*

As a first step, we calculate *analytically* the EM response for (unit-amplitude) plane-wave incidence, and, without losing in generality, we focus on the TM-polarised case (*z*-directed magnetic field). In the fictitious (vacuum) space, no reflection takes place, and the total field coincides with the incident one, which can be written as

$$H_z(x', y') = \exp\left[i\left(k_{x0}x' + k_{y0}y'\right)\right], \qquad (5)$$

where, for a propagating wave with incidence angle $\theta_i$ with respect to the *x*-axis (see figure 1), the *x*- and *y*-domain wave numbers are given by

$$k_{x_0} = k_0 \cos\theta_i, \quad k_{y0} = k_0 \sin\theta_i, \qquad (6)$$

with $k_0 = 2\pi/\lambda_0$ denoting the vacuum wave number (and $\lambda_0$ the corresponding wavelength).

In the standard TO approaches, the field in the physical space would be obtained by straightforward coordinate mapping [via (2)] of the fictitious-space expression in (5). However, in view of the discontinuity of the coordinate transformation in (2) at the interfaces $x=0, x=\pm d$, such mapping is not straightforward here, and special care is needed. Specifically, the physical-space field can be written as a superposition of *forward*- and *backward*-propagating coordinate-mapped [via (2)] plane waves, which can be compactly written as

$$H_z^{(\alpha)}(x,y) = A_\alpha^+ \exp\left[ik_{x\alpha}x'(x,y) + ik_{y\alpha}y'(x,y)\right] + A_\alpha^- \exp\left[-ik_{x\alpha}x'(x,y) + ik_{y\alpha}y'(x,y)\right], \tag{7}$$

with $\alpha = 1, 2$ labelling the bi-layer region [cf. the coordinate mapping in (2)], and $\alpha = 0$ and $\alpha = 3$ labelling the surrounding vacuum regions $x < -d$ and $x > d$, respectively, where an identity transformation $(x' = x, y' = y)$ is assumed. In (7), the amplitude coefficients $A_\alpha^\pm$, $\alpha = 0,1,2,3$, as well as the wave numbers $k_{x\alpha}$ and $k_{y\alpha}$, $\alpha = 1, 2, 3$, need to be computed by enforcing the boundary conditions. First, we note that $A_0^+ = 1$, consistently with the assumed unit-amplitude excitation [cf. (5)], whereas $A_3^- = 0$ in order to fulfil the radiation condition. Next, by enforcing the phase-matching conditions at the interfaces $x = \pm d$ and $x = 0$, we obtain

$$k_{y1} = -ik_{y0}\dot{u}_1(-d), \quad k_{y2} = -ik_{y0}\dot{u}_1(-d)\frac{\dot{u}_2(0)}{\dot{u}_1(0)}, \quad k_{y3} = k_{y0}\frac{\dot{u}_1(-d)\dot{u}_2(0)}{\dot{u}_1(0)\dot{u}_2(d)}, \tag{8}$$

$$k_{x\alpha} = \sqrt{k_0^2 - k_{y\alpha}^2}, \operatorname{Im}(k_{x\alpha}) \geq 0, \quad \alpha = 1, 2, 3, \tag{9}$$

which yield all the unknown wave numbers in (7) as a function of the *y*-domain (vacuum) wave number $k_{y0}$ in (6). Note that, consistently with our original intuition, the complex mapping in (2) transforms *purely imaginary* exponentials in the fictitious space, into *real* ones (with respect to the *x* variable) in the bi-layer region ($\alpha = 1, 2$) of the transformed domain.

Finally, by enforcing the continuity of the tangential field components at $x = \pm d$ and $x = 0$ [with the tangential electric field derived from (7) via the relevant Maxwell's curl equation], we obtain a linear system of six equations, whose (straightforward, but cumbersome) solution yields the remaining unknown amplitude coefficients $A_0^-, A_1^+, A_1^-, A_2^+, A_2^-, A_3^+$. Here, we focus on the coefficient $A_0^-$, which plays the role of the *reflection coefficient*, of direct interest for the subsequent developments, and can be written as

$$A_0^- = \left(\frac{\chi_0 + i\chi_1 \tanh \kappa_{10} + i\chi_2 \tanh \kappa_{20} + \chi_{12} \tanh \kappa_{10} \tanh \kappa_{20}}{\xi_0 + i\xi_1 \tanh \kappa_{10} + i\xi_2 \tanh \kappa_{20} + \xi_{12} \tanh \kappa_{10} \tanh \kappa_{20}}\right) \exp(-2ik_{x0}d), \tag{10}$$

where

$$\kappa_{10} = a_1 k_{x1}\left[u_1(0) - u_1(-d)\right], \quad \kappa_{20} = a_2 k_{x2}\left[u_2(d) - u_2(0)\right], \tag{11}$$

$$\chi_0 = k_{x1}k_{x2}\left[\frac{k_{x3}}{\dot{u}_1(-d)\dot{u}_2(0)} - \frac{k_{x0}}{\dot{u}_1(0)\dot{u}_2(d)}\right], \quad \chi_1 = k_{x2}\left[\frac{k_{x0}k_{x3}}{\dot{u}_2(0)} + \frac{k_{x1}^2}{\dot{u}_1(-d)\dot{u}_1(0)\dot{u}_2(d)}\right],$$

$$\chi_2 = k_{x1}\left[\frac{k_{x0}k_{x3}}{\dot{u}_1(0)} + \frac{k_{x2}^2}{\dot{u}_1(-d)\dot{u}_2(0)\dot{u}_2(d)}\right], \quad \chi_{12} = \frac{k_{x1}^2 k_{x3}}{\dot{u}_1(0)\dot{u}_1(-d)} - \frac{k_{x0}k_{x2}^2}{\dot{u}_2(0)\dot{u}_2(d)}, \tag{12}$$

$$\xi_0 = -k_{x1}k_{x2}\left[\frac{k_{x3}}{\dot{u}_1(-d)\dot{u}_2(0)} + \frac{k_{x0}}{\dot{u}_2(d)\dot{u}_1(0)}\right], \quad \xi_1 = k_{x2}\left[\frac{k_{x0}k_{x3}}{\dot{u}_2(0)} - \frac{k_{x1}^2}{\dot{u}_1(-d)\dot{u}_1(0)\dot{u}_2(d)}\right],$$

$$\xi_2 = k_{x1}\left[\frac{k_{x0}k_{x3}}{\dot{u}_1(0)} - \frac{k_{x2}^2}{\dot{u}_1(-d)\dot{u}_2(0)\dot{u}_2(d)}\right], \quad \xi_{12} = -\frac{k_{x1}^2 k_{x3}}{\dot{u}_1(0)\dot{u}_1(-d)} - \frac{k_{x2}^2 k_{x0}}{\dot{u}_2(0)\dot{u}_2(d)}. \tag{13}$$

*3.2 Total transmission conditions*

Based on the analytical solution above, we can now derive the conditions for *total transmission*, by zeroing the reflection coefficient in (10), viz.,

$$\chi_0 + i\chi_1 \tanh \kappa_{10} + i\chi_2 \tanh \kappa_{20} + \chi_{12} \tanh \kappa_{10} \tanh \kappa_{20} = 0. \tag{14}$$

In general, the above equation needs to be solved *numerically*, and, for a given bi-layer configuration, its solutions [if any, and provided they are not roots of the denominator in (10) too] yield the conditions, in terms of the *y*-domain (vacuum) wave number values $k_{y0}$ in (6) (i.e., the frequency and/or incidence angle), for which the field is totally transmitted through the bi-layer.

In what follows, we focus on a particularly interesting class of solutions, which are independent of $k_{y0}$, and can be studied *analytically*. By inspecting (14), one can observe that, in order for the real part to vanish for any value of $k_{y0}$, the functions $\chi_0$ and $\chi_{12} \tanh \kappa_{10} \tanh \kappa_{20}$ would have to be *linearly dependent*. However, it can be verified that their Wronskian is *generally nonzero*, unless *both* functions vanish[1]. Recalling (11) (and that, by assumption, $a_\alpha \neq 0$ and $\dot{u}_\alpha > 0$), this implies the following *necessary* condition:

$$\chi_0 = \chi_{12} = 0, \quad \forall k_{y0}. \tag{15}$$

From (8) and (9), it is evident that, by enforcing

$$\dot{u}_1(0)\dot{u}_2(d) = \dot{u}_1(-d)\dot{u}_2(0), \tag{16}$$

we obtain $k_{y3} = k_{y0}$ and $k_{x3} = k_{x0}$, which yields [from (12)] $\chi_0 = 0$. Similarly, by enforcing

$$\dot{u}_2(0) = \dot{u}_1(0), \tag{17}$$

we obtain $k_{y1} = k_{y2}$ and $k_{x1} = k_{x2}$, which, together with (16), yields [from (12)] $\chi_{12} = 0$. It can easily be verified that, under these conditions,

$$\chi_1 = \chi_2 = \frac{k_{x1}}{\dot{u}_1(0)}\left[k_{x0}^2 + \frac{k_{x1}^2}{\dot{u}_1(-d)\dot{u}_2(d)}\right], \tag{18}$$

so that the total-transmission condition in (14) can be rewritten as

$$\tanh \kappa_{10} + \tanh \kappa_{20} = 0, \tag{19}$$

which can easily be solved analytically, yielding $\kappa_{20} = -\kappa_{10}$, i.e. [recalling (11)],

$$a_2\left[u_2(d) - u_2(0)\right] = -a_1\left[u_1(0) - u_1(-d)\right]. \tag{20}$$

To sum up, the conditions

$$\begin{cases} \dot{u}_2(0)\dot{u}_1(-d) = \dot{u}_2(d)\dot{u}_1(0), \\ \dot{u}_2(0) = \dot{u}_1(0), \\ a_2\left[u_2(d) - u_2(0)\right] = -a_1\left[u_1(0) - u_1(-d)\right], \end{cases} \tag{21}$$

---

[1] Alternatively, it can be verified that the product of two hyperbolic tangents (of generally nonzero argument) in (14) cannot be exactly represented in terms of the functions $\chi_0$ and $\chi_{12}$ (which do not contain any exponential term).

turn out to be *necessary* and *sufficient* in order to obtain total transmission for *any* value of $k_{y0}$, as it can also be verified that they do not imply the vanishing of the denominator in (10). Note that the conditions in (21) depend on the scaling parameters $a_\alpha$ and on the boundary values of the functions $u_\alpha$ and their derivatives, but are independent of the actual form of these functions as well as of the functions $v_\alpha$. Moreover, while they may be interpreted as *angle-independent* conditions, they implicitly depend on frequency, in view of the unavoidable temporal-dispersion effects in passive SNG media.

Recalling the assumed *positive* character of the derivatives $\dot{u}_\alpha$, which implies that $u_1(0) > u_1(-d)$ and $u_2(d) > u_2(0)$, it readily follows from the third condition in (21) that

$$\text{sgn}(a_1) = -\text{sgn}(a_2), \qquad (22)$$

and thus, depending on the field polarisation (see table 1), we obtain two possible configurations corresponding to the P-ENG/P-MNG and P-MNG/P-ENG pairings.

Under the total-transmission condition, the total phase accumulated by the field transmitted through the bi-layer can be written as

$$\Psi = -k_{y0}\dot{u}_1(-d)\left[-v_1(0) + v_1(-d) + v_2(0) - v_2(d)\right], \qquad (23)$$

and therefore, at normal incidence $(k_{y0} = 0)$ for arbitrary functions $v_\alpha$, or at arbitrary incidence for $v_2(0) - v_2(d) = v_1(0) - v_1(-d)$, the field will undergo a *complete tunnelling* through the bi-layer, without any phase delay. The bi-layer thus behaves as an EM "nullity," in close resemblance with the case of *conjugate matched* (homogeneous, isotropic) ENG-MNG pairs in [20], and its (inhomogeneous, anisotropic) complementary-media generalization in [21,22] (see also the discussion in Sec. 4 below). More in general, an arbitrary aperture field distribution located at a source-plane $x = x_s < -d$ will ideally produce a perfect *virtual* image at the plane $x = x_i \equiv x_s + 2d$, apart from a possible rigid translation of $y_0 = \dot{u}_1(-d)\left[-v_1(0) + v_1(-d) + v_2(0) - v_2(d)\right]$ along the *y*-axis.

## 4. Physical interpretation

*4.1 Special cases*

In order to understand the physical mechanisms underlying the rather general class of *transparent* SNG transformation bi-layers identified by the conditions in (21), it is insightful to start by explicating their relation with the *homogeneous*, *isotropic* ENG-MNG bi-layers investigated in [20], and their further (inhomogeneous, anisotropic) generalizations within the framework on *complementary media* [21,22].

In [20], it was shown that complete tunnelling (with zero phase delay) could be achieved by pairing two slabs (of thickness $d_1$ and $d_2$) made of homogenous, isotropic ENG and MNG materials. In particular, under the *conjugate-matching* condition:

$$\varepsilon_2 = -\varepsilon_1, \quad \mu_2 = -\mu_1, \quad d_2 = d_1, \qquad (24)$$

such tunnelling effects would occur for *any* value of the incident wave number, as long as the above condition (24) remains satisfied. Such conditions were further generalised in [21,22] to the case of *anisotropic*, *inhomogeneous* (possibly lossy) materials satisfying the complementary-media condition, which, in our chosen reference system, can be expressed as

$$\underline{\underline{\varepsilon}}_2 = \begin{bmatrix} -\varepsilon_{1xx} & \varepsilon_{1xy} & \varepsilon_{1xz} \\ \varepsilon_{1yx} & -\varepsilon_{1yy} & -\varepsilon_{1yz} \\ \varepsilon_{1zx} & -\varepsilon_{1zy} & -\varepsilon_{1zz} \end{bmatrix}, \quad \underline{\underline{\mu}}_2 = \begin{bmatrix} -\mu_{1xx} & \mu_{1xy} & \mu_{1xz} \\ \mu_{1yx} & -\mu_{1yy} & -\mu_{1yz} \\ \mu_{1zx} & -\mu_{1zy} & -\mu_{1zz} \end{bmatrix}. \qquad (25)$$

It can easily be verified that our general class in (21) reduces to the complementary-media case in (25) for

$$\begin{cases} a_2 = -a_1, \\ u_2(x) = -u_1(-x), \\ v_2(x) = v_1(-x), \end{cases} \quad (26)$$

which, in turn, yields the homogeneous, isotropic case in (24) for $u_2(x) = x$ and $v_2(x) = 0$.

In this framework, we point out that some seeming restrictions in our class [e.g., the fact that $\underline{\underline{\varepsilon}}_\alpha = \underline{\underline{\mu}}_\alpha$, cf. (1)] are only due to our original choice of transforming a *vacuum* fictitious space, and can accordingly be overcome by considering a more general filling (not necessarily homogeneous and isotropic, and possibly DNG).

*4.2 Complex-mapping-induced surface modes*

While the complementary-media arguments above provide an interesting interpretation of some special cases, the conditions in (21) define a broad class of configurations that do not generally satisfy the symmetry conditions in (24) or (25).

A deeper physical insight can be gained by looking at the analytical structure of the field inside the bi-layer region. As anticipated, one intuitively expects the in-plane purely-imaginary character of the coordinate transformation in (2) to map a *propagating* field [in the vacuum fictitious space, cf. (5)] into an *evanescent* one in the P-SNG transformed domain. However, when pairing two conjugate SNG transformation slabs, an *exponentially-growing* behaviour may also occur. Specifically, by particularising the general expression in (7) under the total-transmission conditions in (21), we obtain

$$H_z^{(\alpha)}(x,y) = \left\{ \cosh\left[\kappa_\alpha(x)\right] \pm i \frac{k_{x0} \dot{u}_1(-d)}{k_{x1}} \sinh\left[\kappa_\alpha(x)\right] \right\} \\ \times \exp\left[ -i k_{x0} d + i k_{y0} y \frac{\dot{u}_1(-d)}{\dot{u}_\alpha(x)} \right], \quad \alpha = 1, 2, \quad (27)$$

where

$$\kappa_\alpha(x) = \pm a_\alpha k_{x\alpha} \left[ u_\alpha(x) - u_\alpha(\mp d) \right]. \quad (28)$$

We note that the field exhibits unit amplitude at the interfaces $x = \pm d$ (as expected, in order to match the incident and totally-transmitted fields), while its magnitude inside the bi-layer region is *always* greater than unity, exponentially growing towards the interface $x = 0$, where it reaches its peak. Thus, the complex-mapping in (2), under the total-transmission conditions in (21), transforms a *propagating* (constant-intensity) plane-wave field in the vacuum fictitious domain [see figure 2(a)] into a *surface mode* localised at the interface $x = 0$ separating the two opposite-signed P-SNG slabs in the physical space [see figure 2(b)].

In the complementary-media case [20-21], the presence of surface modes has been observed and associated with tunnelling phenomena. Our complex-mapping arguments above provide an alternative TO-based interpretation, and extend the association between surface modes and tunnelling phenomena beyond the complementary-media scenario.

## 5. Representative numerical results

As a representative example, we consider a configuration featuring a *homogeneous*, *isotropic* MNG slab $\left( \tilde{\varepsilon}_{1\xi} = \tilde{\varepsilon}_{1\upsilon} = 1, \tilde{\mu}_{1z} = -1 \right)$, i.e.,

$$a_1 = 1, \quad u_1(x) = x, \quad v_1(x) = 0, \quad (29)$$

paired with an *inhomogeneous*, *anisotropic* P-ENG transformation slab $\left(\tilde{\varepsilon}_{2\xi}<0, \tilde{\varepsilon}_{2\upsilon}<0, \tilde{\mu}_{2z}=1\right)$, obtained via

$$a_2 = -1, \quad u_2(x) = d\left[\frac{x}{d} - 3\left(\frac{x}{d}\right)^2 + 6\left(\frac{x}{d}\right)^3 - 3\left(\frac{x}{d}\right)^4\right], \tag{30}$$

where the polynomial coefficients in $u_2$ have been chosen so as to satisfy the total-transmission conditions in (21), as well as to ensure that $\dot{u}_2(x)>0$ within the interval $0<x<d$. Figure 3 illustrates the distributions of the relative-permittivity principal components of the P-ENG transformation slab, from which the inhomogeneous, anisotropic (and, obviously, *negative*) character is evident. We highlight that the bi-layer configuration defined by (29) and (30) does *not* satisfy the symmetry conditions in (24) or (25), and thus its EM response cannot be explained within the complementary-media framework.

Figure 4 illustrates the EM response to a normally-incident $(\theta_i = 0)$, unit-amplitude plane-wave illumination, for a thickness $d = \lambda_0/3$. Specifically, figures 4(a) and 4(b) show the field magnitude (with stream lines illustrating the local energy flux) and phase maps, respectively, whereas figure 4(c) shows a longitudinal cut of the field magnitude. One can readily observe the predicted localization effects (surface mode) at the interface $x=0$ (better quantified in the longitudinal cut), as well as the perfect matching between the phase profiles at the input $(x=-d)$ and output $(x=d)$ interfaces, indicative of the *zero phase-delay* induced by the bi-layer.

Qualitatively similar considerations hold for the case of oblique incidence $(\theta_i = 15°)$, illustrated in figure 5, with an expected increased complexity in the internal field structure and energy flux.

We also investigated the effects of the unavoidable material losses in the above illustrated tunnelling phenomena (see [20,22] for the complementary-media case). With reference to the parameter configurations in figures 4 and 5, figure 6 shows the transmission coefficient magnitude as a function of the material loss-tangent $\tan\delta$ (assumed identical for the two constituents of the bi-layer). The lossy configuration was obtained as a perturbation of the lossless one above, by multiplying all the negative constitutive parameters $\left(\tilde{\mu}_{1z}, \tilde{\varepsilon}_{2\xi}, \tilde{\varepsilon}_{2\upsilon}\right)$ by a factor $(1-i\tan\delta)$, so as to ensure the *positivity* of their imaginary parts, and hence the *passivity* of the media. As it can be observed, the tunnelling effects turn out to be rather robust up to moderate (loss-tangent=0.01) levels of losses. Being inherently a resonant phenomenon, however, significant metamaterial losses may significantly affect the tunnelling, as confirmed in Fig. 6. Analogous to the case of complementary SNG pairs [20], the loss effect is predicted to be more sensitive to thicker complementary layers.

## 6. Conclusions

In this paper, we have presented a TO-based interpretation of tunnelling phenomena in bi-layers made of paired P-ENG/P-MNG media. Our proposed interpretation, which relies on a *complex* coordinate transformation, includes the (homogeneous, isotropic) scenario in [20] and its (inhomogeneous, anisotropic) complementary-media extensions in [21,22], and suggests other interesting generalizations, thereby providing further evidence of the versatility and *unifying* character of TO.

Current and future research are aimed at the exploration of different TO-inspired scenarios, such as bi-layers made of P-SNG and DPS media, for which interesting tunnelling effects may occur at given frequencies and incidence angles.

**Table 1.** Possible characters of the SNG transformation media in (3), depending on the field polarisation and on the scaling parameter sign.

| Polarisation | Scaling parameter | |
|---|---|---|
| | $a_\alpha > 0$ | $a_\alpha < 0$ |
| TE | $\tilde{\varepsilon}_{\alpha z} < 0, \tilde{\mu}_{\alpha \xi} > 0, \tilde{\mu}_{\alpha \upsilon} > 0$ (ENG) | $\tilde{\varepsilon}_{\alpha z} > 0, \tilde{\mu}_{\alpha \xi} < 0, \tilde{\mu}_{\alpha \upsilon} < 0$ (MNG) |
| TM | $\tilde{\varepsilon}_{\alpha \xi} > 0, \tilde{\varepsilon}_{\alpha \upsilon} > 0, \tilde{\mu}_{\alpha z} < 0$ (MNG) | $\tilde{\varepsilon}_{\alpha \xi} < 0, \tilde{\varepsilon}_{\alpha \upsilon} < 0, \tilde{\mu}_{\alpha z} > 0$ (ENG) |

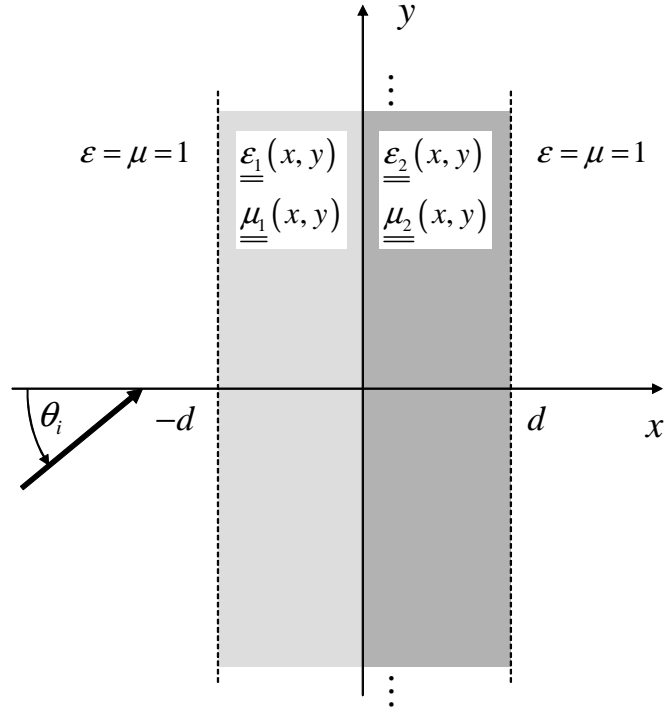

**Figure 1.** Geometry of the problem. Two transformation-slabs (each of thickness $d$, and infinitely long in the $y$ and $z$ directions) with constitutive tensors given by (1) are paired in a vacuum space. Also shown is the direction and incidence angle $\theta_i$ pertaining to the plane-wave excitation case considered.

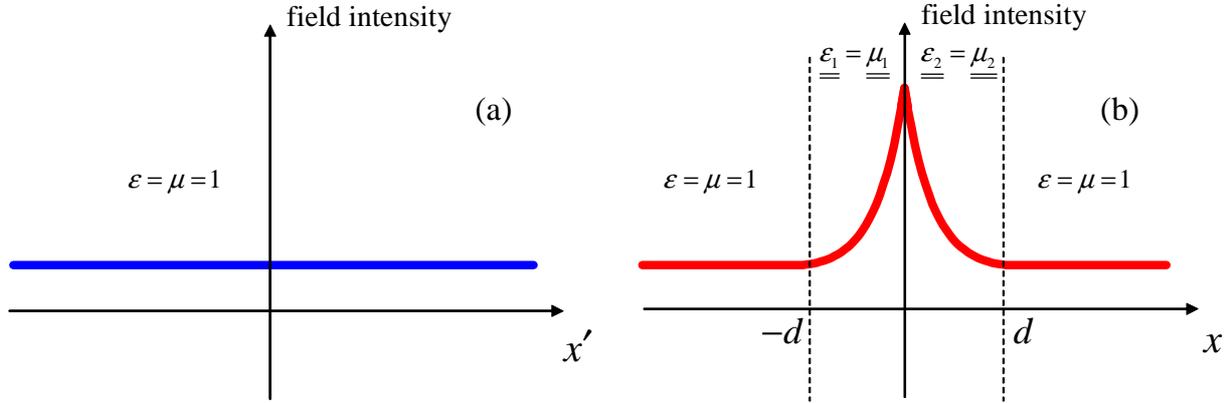

**Figure 2.** (Colour online) Schematic illustration of the effects of the complex coordinate transformation in (2) which, under the total-transmission conditions in (21), transforms a *propagating* (constant-intensity) plane-wave field (a) in the vacuum fictitious space into a *localised surface mode* (b) in the physical space.

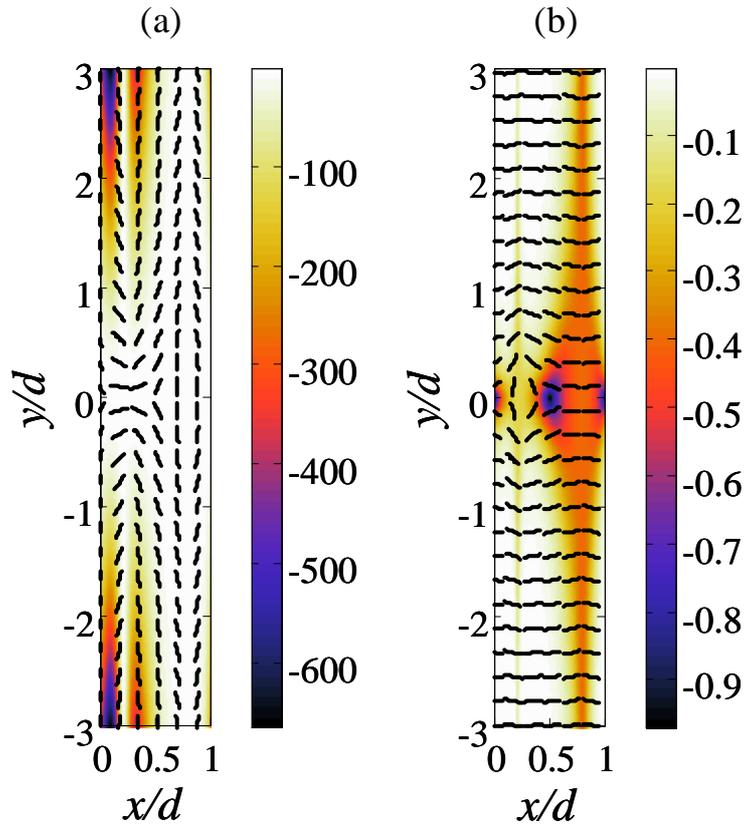

**Figure 3.** (Colour online) Relative-permittivity maps (in-plane components) pertaining to the P-ENG transformation slab in (30), shown in the principal reference system. As a reference, the principal axes directions $\xi$ and $\upsilon$ are shown as short segments, in (a) and (b), respectively.

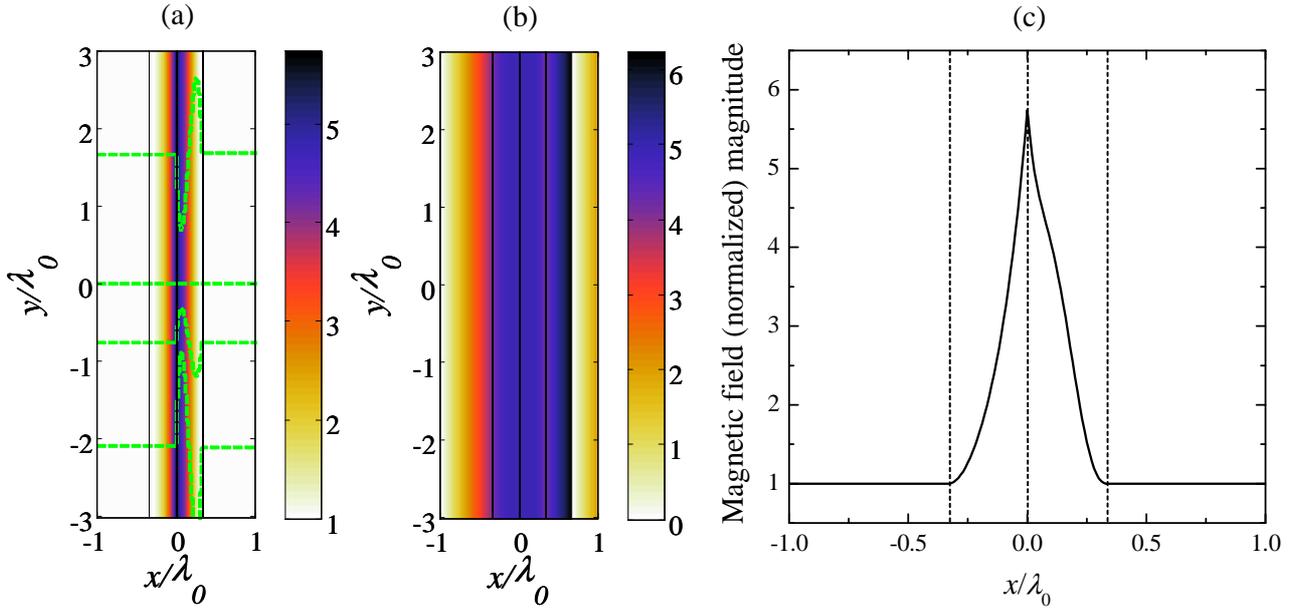

**Figure 4.** (Colour online) EM response, for TM-polarised normally-incident $(\theta_i = 0)$ plane-wave excitation, of a bi-layer obtained by pairing a homogeneous MNG slab [cf. (29)] with the P-ENG transformation slab in figure 3, for $d = \lambda_0/3$. (a), (b): Magnetic field magnitude and phase (radian) maps, respectively, of the magnetic field component, with some representative (green-dashed) streamlines indicating the local energy flux (i.e., Poynting vector). (c): Field-magnitude longitudinal cut (at $y = 0$).

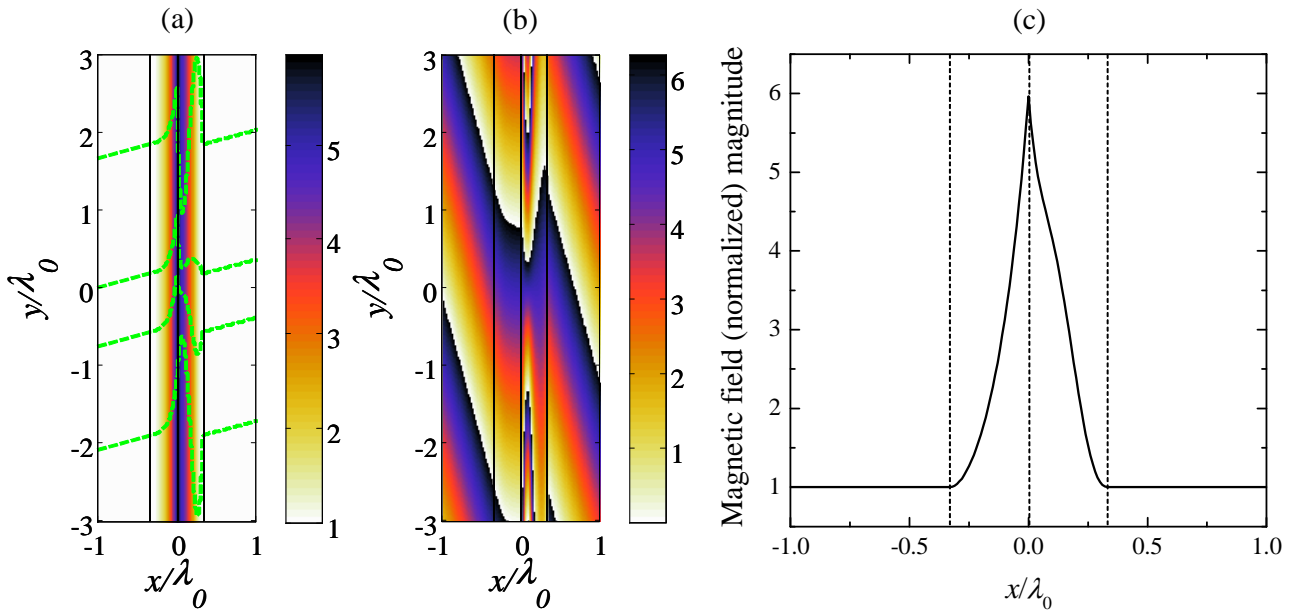

**Figure 5.** (Colour online) As in figure 4, but for oblique incidence $(\theta_i = 15°)$.

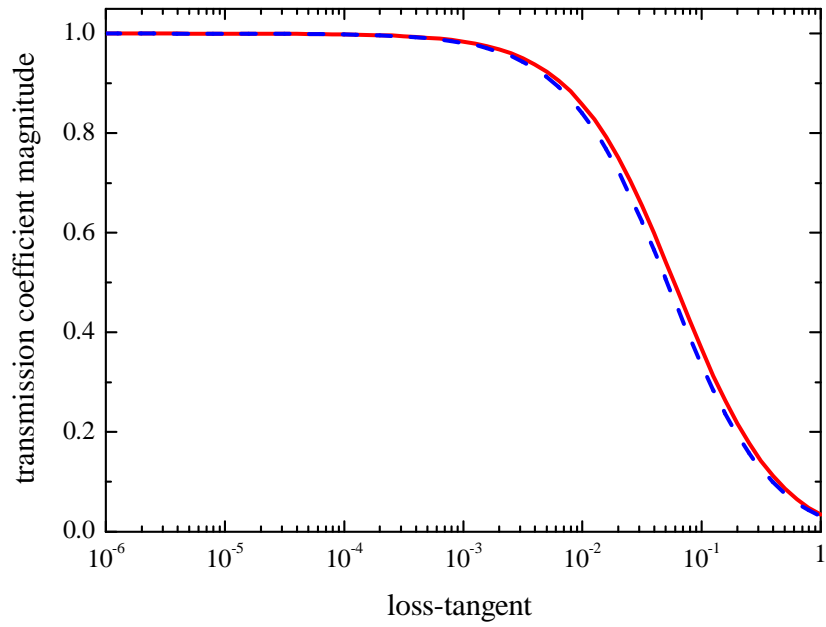

**Figure 6.** (Colour online) Parameters as in figures 4 and 5. Transmission coefficient magnitude as a function of the material loss-tangent (assumed identical for the two constituents of the bi-layer), for normal (red-solid curve) and oblique ($\theta_i = 15°$, blue-dashed curve) incidence.